\documentclass{article}

\usepackage{microtype}
\usepackage{graphicx}
\usepackage{subfigure}
\usepackage{booktabs} 

\usepackage{multirow}
\usepackage{tabularx}
\usepackage{kotex}
\usepackage{comment}
\usepackage{soul}

\usepackage{hyperref}

\usepackage[accepted]{icml2020}

\renewcommand{\footnotesize}{\fontsize{7pt}{11pt}\selectfont}

\icmltitlerunning{Musical Word Embedding: Bridging the Gap between Listening Contexts and Music}

\begin{document}

\twocolumn[
\icmltitle{Musical Word Embedding: Bridging the Gap between Listening Contexts and Music}



\icmlsetsymbol{equal}{*}

\begin{icmlauthorlist}
\icmlauthor{Seungheon Doh}{kaist}
\icmlauthor{Jongpil Lee}{kaist}
\icmlauthor{Tae Hong Park}{nyu}
\icmlauthor{Juhan Nam}{kaist}
\end{icmlauthorlist}

\icmlaffiliation{kaist}{Graduate School of Culture Technology, Korea Advanced Institute of Science and Technology (KAIST), Daejeon, South Korea}
\icmlaffiliation{nyu}{Steinhardt School of Culture, Education, and Human Development, New York University, New York, United States}

\icmlcorrespondingauthor{Juhan Nam}{juhan.nam@kaist.ac.kr}

\icmlkeywords{music word2vec}


\vskip 0.3in
]



\printAffiliationsAndNotice{}  

\begin{abstract}
Word embedding pioneered by Mikolov et al. is a staple technique for word representations in natural language processing (NLP) research which has also found popularity in music information retrieval tasks. Depending on the type of text data for word embedding, however, vocabulary size and the degree of musical pertinence can significantly vary. In this work, we (1) train the distributed representation of words using combinations of both general text data and music-specific data and (2) evaluate the system in terms of how they associate listening contexts with musical compositions. 
\vspace{-3mm}
\end{abstract}

\vspace{-2mm}
\section{Introduction}
\vspace{-2mm}
Music listeners typically rely on a combination of listening contexts to find music including elements of mood, theme, time of day, location and activity. This scenario can be handled by defining a dictionary of contextual terms and directly associating them with music as a class label \cite{yan2015improving, ibrahim2020audio}. However, such a music tagging approach (i.e., multi-label classification) is severely limited in considering contextual expression complexities that listeners can use from a natural language perspective. For example, a listener may use `club' to search for electronic dance music, and unless a model is trained with this specific word, it is not possible to consider the word as a query string. This issue has been addressed by representing tag words with embedding vectors and associating them with music in several different settings such as zero-shot learning \cite{choi2019zero}, query-by-blending \cite{watanabe2019query} and multi-task music representation learning \cite{schindler2019multi}. The aforementioned approaches were based on system training utilizing word embedding with either general text (e.g., Wikipedia or Gigaword) or music-specific corpus (e.g., tags, lyrics, artist IDs, track IDs). What is noteworthy here is that the general text training approach is limited in reflecting "musical" dimensions, whereas music-specific corpus limits incorporation of listening contexts which are not directly related to music while simultaneously suffering from small vocabulary size. 
In this work, we investigate various word embedding spaces trained with  combinations of general and music-specific text data to bridge the gap between listening contexts and music. 


\vspace{-3mm}
\section{Datasets and Method} 
\vspace{-2mm}
We conducted our research using the latest \textit{Wikipedia} dump\footnote{\url{https://dumps.wikimedia.org/enwiki/20200601/}} for general text data and a hybrid music corpus for music-specific text data. The music corpus is composed of \textit{Amazon} album review, \textit{AllMusic} tags\footnote{\url{https://www.allmusic.com}}, and artist/track IDs. The \textit{Amazon} album review data contain consumer opinions about the music \cite{He2016}, which was obtained from the MuMu dataset \footnote{\url{https://www.upf.edu/web/mtg/mumu}} \cite{oramas2017multi}. The \textit{Allmusic} dataset includes music tags (genre, style) and context tags (mood and theme) \cite{schindler2019multi}. The artist/track IDs were obtained from the MSD dataset \cite{bertin2011million}. The IDs are also regarded as a unique word associated with the corresponding music \cite{watanabe2019query}. 
We used Word2Vec based on Continuous Bag of Words (CBOW) to learn word embedding \cite{mikolov2013efficient}. For music corpus, we clustered review texts, tags, and artist/track IDs for each music track using MSD track\_id \footnote{\url{http://millionsongdataset.com/}} and MusicBrainz id \footnote{\url{https://musicbrainz.org/}} and also took the context window within the cluster. Additionally, we shuffled words within clusters to address data augmentation. Although this method broke review sentence order, it improved capture of word co-occurrences in the hybrid set with greater spread.


\begin{table*}[!t]
\begin{center}
\vspace{-4mm}
\caption{Compare ranking evaluation metric between 7 embedding spaces.} 
\label{tab:1}
\begin{tiny}
\begin{sc}
\resizebox{0.9\textwidth}{!}{
\begin{tabular}{ccccccccc}
\toprule
\multirow{2}{*}[-0.7ex]{Corpus} & \multirow{2}{*}[-0.7ex]{Size} & \multirow{2}{*}[-0.7ex]{\begin{tabular}[c]{@{}c@{}}Unique \\ Word\end{tabular}} & \multirow{2}{*}[-0.7ex]{\begin{tabular}[c]{@{}c@{}}Unique \\ Track\end{tabular}} & \multirow{2}{*}[-0.7ex]{\begin{tabular}[c]{@{}c@{}}Unique \\ Artist\end{tabular}} & \multicolumn{2}{c}{AllMusic (Seen)} & \multicolumn{2}{c}{LastFm (Unseen)} \\ \cmidrule(lr){6-7} \cmidrule(lr){8-9}
 &  &  &  &  & spearmanr & nDCG@30 & spearmanr & nDCG@30 \\ \midrule
{[}AllMusic Tags   + Amazon Music Reviews{]} (Augmented) + Wikipedia & 1.98B & 11,622,471 & 521,778 & 28,330 & 0.194 & 0.327 & 0.312 & 0.591 \\ \cmidrule{1-9}
AllMusic Tags + Amazon Music Reviews +   Wikipedia & 1.8B & 11,622,471 & 521,778 & 28,330 & 0.187 & 0.233 & 0.226 & 0.548 \\ \cmidrule{1-9}
AllMusic Tags   + Wikipedia & 1.76B & 11,163,229 & 507,435 & 25,203 & 0.157 & 0.215 & 0.183 & 0.526 \\ \midrule
{[}AllMusic Tags   + Amazon Music Reviews{]} (Augmented) & 0.27B & 664,163 & 521,778 & 28,330 & \textbf{0.267} & \textbf{0.339} & \textbf{0.407} & \textbf{0.626} \\ \cmidrule{1-9}
AllMusic Tags + Amazon Music Reviews & 45.3m & 664,163 & 521,778 & 28,330 & 0.187 & 0.232 & 0.358 & 0.600 \\ \cmidrule{1-9}
AllMusic Tags & 7.1m & 1,401 & 507,435 & 25,203 & 0.252 & 0.242 &  &  \\ \midrule
Wikipedia & 1.75B & 11,163,055 & 0 & 0 & 0.098 & 0.167 & 0.162 & 0.551
\\\bottomrule
\end{tabular}
}
\end{sc}
\end{tiny}
\end{center}
\vspace{-3mm}
\end{table*}

\begin{figure*}[!t]
\begin{center}
\centerline{\includegraphics[width=0.95\textwidth]{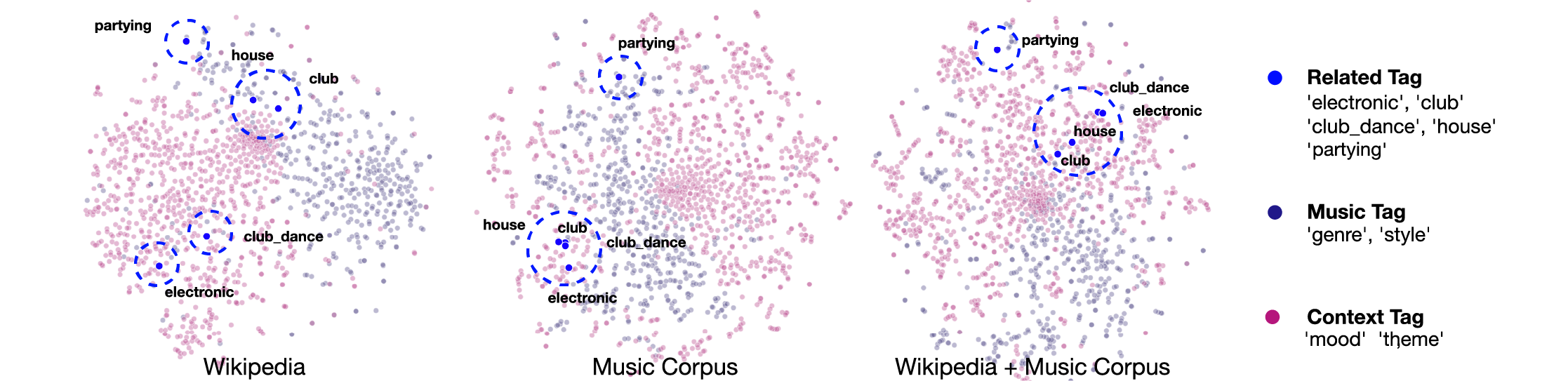}}
\vspace{-2mm}
\caption{Tag visualization of different type of embedding using t-SNE. The music corpus includes \textit{AllMusic} Tags, \textit{Amazon} music reviews, MSD Artist/Track IDs.  
}
\label{figure:fig1}
\end{center}
\vspace{-8mm}
\end{figure*}


\vspace{-3mm}
\section{Experiments}
\vspace{-2mm}

We trained the word embedding model with vector size 100, window size 15, and five iterations. To test word embedding, we used test tags from two different datasets and thus, different characteristics. \textit{Allmusic} was used to enable a balanced distribution of music terms and context terms consisting of 1,401 genre/style music tags and mood/theme context tags. This dataset was also used for training (seen) as part of the music corpus. The \textit{last.fm} dataset with genre, mood, and eras tags was also used. We selected the top 100 tags with maximal occurrence frequency. The latter dataset which focuses on music terms was not included in the training (unseen) phase. To measure word-to-word similarity performance of the proposed word embedding system, we employed a co-occurrence of tags scheme for ground truth creation. We then measured spearman's rank correlation and normalized discounted cumulative gain at \textit{k} (nDCG@k) between ground-truth co-occurrence and word-to-word similarity of word embeddings. For the nDCG evaluation, we use the top \textit{k} retrieved words ($k=30$). 

\vspace{-3mm}
\section{Results and Discussion} 
\vspace{-2mm}
Table \ref{tab:1} shows performance results, size of the training corpus, unique words, unique tracks, and unique artists of each method. The results show that the two word embeddings including music corpus significantly outperform the model trained with \textit{Wikipedia} only. This is expected as the test sets were based on music tag datasets. Between music corpus only and music corpus with Wikipedia, the result depended on how many music terms and context terms are balanced in the test sets. When music terms are concentrated (\textit{last.fm} tags), word embedding trained with music corpus only outperformed that with both music corpus and Wikipedia. However, in the balanced case (\textit{AllMusic} tags), word embedding trained with both music corpus and Wikipedia resulted in improved performances. Table \ref{tab:1} also shows that the augmented music corpus achieved notable high performance results. This suggests that the proposed data augmentation is beneficial when the order of words is not important. 
The t-SNE plot in Figure \ref{figure:fig1} provides a more intuitive visualization of our research results. Here, we used two music genre terms `electronic' and `house' and three listening context terms `club', `club\_dance', and `partying' as relevant words. In \textit{Wikipedia}, the gaps between terms are significant with only `house' and `club' in close proximity. In the music corpus, the two genre terms and `club' and `club\_dance' are tightly clustered while `partying' is significantly beyond the cluster centroid. In the music corpus with \textit{Wikipedia}, while the context term `partying' is still outside of the cluster containing all of the other terms, it is substantially closer than the music corpus example. This indicates that using both general and music-specific data has the potential of capturing a more balanced correlation between music and listening context (for examples of music retrieval tasks using context words, please refer to \footnote{\url{https://dohppak.github.io/MusicWordVec}}).

\vspace{-3mm}
\section{Future Work}
\vspace{-2mm}
Our current plan is to expand on findings as reported in this paper and build a set of user-annotated word-to-word similarity pairs to directly measure the relationship between general words, music contexts, and music tracks. We also plan to additionally use the musical word embedding system from trained word embedding as a prototype vector for each music track in the context of audio-based music regression \cite{van2013deep}, classification, and metric learning \cite{choi2019zero} settings. This will allow us to construct a more nuanced audio embedding system as conventional music classification is in the order hundreds of labels as class prototypes, while the proposed approach allows for half a million of track prototypes that are strongly reflective of millions of music context terms. 

\nocite{langley00}

\bibliography{ref}

\begin{thebibliography}{10}
\providecommand{\natexlab}[1]{#1}
\providecommand{\url}[1]{\texttt{#1}}
\expandafter\ifx\csname urlstyle\endcsname\relax
  \providecommand{\doi}[1]{doi: #1}\else
  \providecommand{\doi}{doi: \begingroup \urlstyle{rm}\Url}\fi

\bibitem[Bertin-Mahieux et~al.(2011)Bertin-Mahieux, Ellis, Whitman, and
  Lamere]{bertin2011million}
Bertin-Mahieux, T., Ellis, D.~P., Whitman, B., and Lamere, P.
\newblock The million song dataset.
\newblock In \emph{Proc. International Society for Music Information Retrieval
  Conference (ISMIR)}, 2011.

\bibitem[Choi et~al.(2019)Choi, Lee, Park, and Nam]{choi2019zero}
Choi, J., Lee, J., Park, J., and Nam, J.
\newblock Zero-shot learning for audio-based music classification and tagging.
\newblock In \emph{Proc. International Society for Music Information Retrieval
  Conference (ISMIR)}, pp.\  67--74, 2019.

\bibitem[He \& McAuley(2016)He and McAuley]{He2016}
He, R. and McAuley, J.
\newblock Ups and downs: Modeling the visual evolution of fashion trends with
  one-class collaborative filtering.
\newblock In \emph{Proceedings of the 25th International Conference on World
  Wide Web}, pp.\  507–517, 2016.

\bibitem[Ibrahim et~al.(2020)Ibrahim, Royo-Letelier, Epure, Peeters, and
  Richard]{ibrahim2020audio}
Ibrahim, K.~M., Royo-Letelier, J., Epure, E.~V., Peeters, G., and Richard, G.
\newblock Audio-based auto-tagging with contextual tags for music.
\newblock In \emph{Proc. International Conference on Acoustics, Speech and
  Signal Processing (ICASSP)}, pp.\  16--20. IEEE, 2020.

\bibitem[Mikolov et~al.(2013)Mikolov, Chen, Corrado, and
  Dean]{mikolov2013efficient}
Mikolov, T., Chen, K., Corrado, G., and Dean, J.
\newblock Efficient estimation of word representations in vector space.
\newblock In \emph{Proc. International Conference on Learning
  Representations,Workshop Track Proceedings,{ICLR}}, 2013.

\bibitem[Oramas et~al.(2017)Oramas, Nieto, Barbieri, and
  Serra]{oramas2017multi}
Oramas, S., Nieto, O., Barbieri, F., and Serra, X.
\newblock Multi-label music genre classification from audio, text, and images
  using deep features.
\newblock In \emph{Proc. International Society for Music Information Retrieval
  Conference (ISMIR)}, pp.\  23--30, 2017.

\bibitem[Schindler \& Knees(2019)Schindler and Knees]{schindler2019multi}
Schindler, A. and Knees, P.
\newblock Multi-task music representation learning from multi-label embeddings.
\newblock In \emph{Proc. International Conference on Content-Based Multimedia
  Indexing (CBMI)}, pp.\  1--6. IEEE, 2019.

\bibitem[Van~den Oord et~al.(2013)Van~den Oord, Dieleman, and
  Schrauwen]{van2013deep}
Van~den Oord, A., Dieleman, S., and Schrauwen, B.
\newblock Deep content-based music recommendation.
\newblock In \emph{Advances in neural information processing systems}, pp.\
  2643--2651, 2013.

\bibitem[Watanabe \& Goto(2019)Watanabe and Goto]{watanabe2019query}
Watanabe, K. and Goto, M.
\newblock Query-by-blending: a music exploration system blending latent vector
  representations of lyric word, song audio, and artist.
\newblock In \emph{Proc. International Society for Music Information Retrieval
  Conference (ISMIR)}, pp.\  144--151, 2019.

\bibitem[Yan et~al.(2015)Yan, Ding, Yin, and Lv]{yan2015improving}
Yan, Q., Ding, C., Yin, J., and Lv, Y.
\newblock Improving music auto-tagging with trigger-based context model.
\newblock In \emph{Proc. International Conference on Acoustics, Speech and
  Signal Processing (ICASSP)}, pp.\  434--438. IEEE, 2015.

\end{thebibliography}
\bibliographystyle{icml2020}





\end{document}